\documentclass[aps,showpacs, nofootinbib]{revtex4}

\usepackage{graphicx}

\newcommand{\be}{\begin{equation}}
\newcommand{\ee}{\end{equation}}
\newcommand{\ben}{\begin{eqnarray}}
\newcommand{\een}{\end{eqnarray}}

\newcommand{\om}{{\omega}}

\newcommand{\tc}{{\tilde c}}

\newcommand{\cA}{{\cal A}}
\newcommand{\cB}{{\cal B}}

\newcommand{\na}{\nabla}

\newcommand{\tpsi}{\tilde \psi}

\newcommand{\tom}{{\tilde \omega}}

\newcommand{\tA}{{\tilde A}}

\newcommand{\tM}{\tilde M}


\begin{document}

\title{Influence of a Brane Tension on Phantom and 
Massive Scalar Field Emission} 

\author{Marek Rogatko and Agnieszka Szyp\l owska}
\affiliation{Institute of Physics \protect \\
Maria Curie-Sklodowska University \protect \\
20-031 Lublin, pl.~Marii Curie-Sklodowskiej 1, Poland \protect \\
rogat@kft.umcs.lublin.pl\protect \\
marek.rogatko@poczta.umcs.lublin.pl}

\date{\today}

\begin{abstract}
We elaborate the signature of the extra dimensions and brane tension in the process
of phantom and massive scalar emission in the spacetime of $(4 + n)$-dimensional tense
brane black hole. Absorption cross section, luminosity of Hawking radiation
and cross section in the low-energy approximation were found. We envisage that parameter
connected with the existence of a brane imprints its role in
the Hawking radiation of the considered fields.

\end{abstract}

\maketitle

\section{Introduction}
Theories with extra dimensions attract much attention due to the fact of resolution of the long standing
question about hierarchy problem. This problem arises between the scale of gravity and the other 
fundamental interactions. Braneworlds models \cite{ark98,ran99} point the way out of it by lowering the
fundamental scale of gravity down to order of TeV. Another interesting possibility that appears in the content
of the so-called TeV-gravity is the possibility of production mini black holes through high-energy collisions.
The underlying process opens for us
the window to study gravity at small distances as well as to research physics of extra dimensions.
Black holes created in high-energy collisions were expected to evaporate through Hawking radiation
both in the bulk (emission of gravitons and scalar fields) and on the brane (by emission of fermions and gauge bosons).
It is expected that the newly born black hole will undergo a number of phases.
Namely, {\it balding phase} when black hole will emit mainly gravitational radiation. Then, it will be
{\it spin-down phase} during which black hole will loose its angular momentum through emission of Hawking radiation.
Next phase is called {\it Schwarzschild phase} in which black hole will loose its actual mass by Hawking radiation.
In {\it Planck phase} quantum gravity theory is needed to study its behaviour.\\
One of the striking feature arising due to the TeV range of energy of CERN Large Hadron Collider will be the fact that it can 
a mini black hole factory. 
All these motivate the recent and future extensive studies on this breath taking subject.
\par
On the other hand,
the {\it no-hair theorem} for $n$-dimensional static black holes was quite well justified \cite{nhair},
as well as the mechanism of decaying black hole hair in higher dimensions was widely studied (see, e.g.,
\cite{dec}). As far as the particle emission from multidimensional black hole is concerned it also has its own long history.
Namely, in Ref.\cite{kan03a} massless scalar emission were studied in the context
of $(4 + n)$-dimensional Schwarzschild black hole, while the low-energy absorption cross section for massive scalar 
and Dirac fermion in the spacetime in question was studied in Ref.\cite{jun04}.
The studies were broadened
to the case of massless spinor and gauge particles \cite{kan03b}. 
Then, the emitted radiation from higher-dimensional black holes were studied both analytically and numerically
(see, e.g., Refs.\cite{emission} for a non-exhaustive sampling of this widely treated subject).
Graviton emission in the bulk from a higher dimensional Schwarzschild
black hole was elaborated in Refs.\cite{cre06}. It was established among all, that the low-energy emission rate
decreases with the number of extra-dimensions as was previously found for the case of bulk
massless scalar field.
\par
The complexity of the aforementioned problem in the background of a rotating $(4 + n)$-dimensional black hole
was revealed in Refs.\cite{cre07} (see also Refs.\cite{fro03} devoted to the particular case of five-dimensional
rotating black hole).
\par
Recently, the metric of four-dimensional acoustic black hole was extended to the case of higher 
dimensions in order to study the emission of massless scalar fields in such spacetime \cite{ge07}.
\par
The absorption probability of massless scalar field were also studied in the rotating G\"odel
black hole in minimal five-dimensional gauged supergravity \cite{che08} as well as in the spacetime of rotating
Kaluza-Klein black hole with squashed horizon \cite{che08b}. Hawking radiation of $(4 + n)$-dimensional
Schwarzschild black hole imbeded in de Sitter spacetime was investigated in Ref.\cite{wu08}.
On the other hand, emission
of massless scalar fields in the background of $n$-dimensional static black hole
surrounded by quintessence was elaborated in Ref.\cite{che08c}, while Hawking radiation of phantom field
in Kerr background was analyzed in Ref.\cite{chejing08}.

\par 
Most examinations of extra-dimensions
black hole and their physics were devoted to the zero brane tension case. But in principle
finite brane tension may modify the physics of a tensional brane black hole. 
The nonzero tension on the brane can curve the brane as well as the bulk.
A tense brane black hole is locally a higher-dimensional Schwarzschild solution \cite{kal06}
threatened by a tensional brane. This caused that a deficit angle appeared in the $(n + 2)$-dimensional
unit sphere line element.
Therefore more attention should be confined to studies of the aforementioned black objects.
Some hints to achieve these goals were conducted. Namely,
studies of massless fermion excitation on a tensional three-brane were carried in Ref.\cite{cho08},
while the late-time behaviour of massive scalar hair in the background of $n$-dimensional tense brane black hole was 
studied in \cite{rog08}.
Emissions of massless scalar fields into the bulk from six-dimensional rotating black hole pierced by a three-brane
was studied in \cite{kob08}. On the other hand, Ref.\cite{dai07} was devoted to the numerical
studies of evaporation of massless scalar,
vector and graviton fields in the background of a six-dimensional tense brane black hole. 
Growing interests in codimensional-2 braneworlds lead also to modify
the gravitational action by implementing Gauss-Bonnet term or to consider black hole solutions
on a thin three-brane of codimension-2 \cite{cua08,cua08b}.
\par 
Nevertheless, to our knowledge, up to now, there was no analytical studies of massive scalar and phantom field emission from 
a tense brane black hole in arbitrary dimension. 
In the present work we shall 
generalize the previous studies taking into account non zero brane tension as well as considering
both massive scalar and phantom fields in the spacetime of a tense brane black hole of $(n + 4)$-dimension.
In Sec.II we focus on the low-energy regime and solve analytically Eqs. of motion for 
the adequate fields by means of 
the {\it matching technique} combining the far field and near event horizon solutions.
Then, we find the analytical expression for the absorption cross section, luminosity of
Hawking radiation as well as the low-energy cross section. Our analytical consideration are supplemented by 
plots expressing the dependence of the absorption cross section on various parameters of the considered spacetime.
Sec.III will be devoted to our conclusions.

\section{Greybody Factor in the Low-energy Regime}
Greybody factors are of great importance for they enable us to study the near horizon structure
of black holes. They are also important from the experimental point of view due to the fact that they
modify the spectrum in the region of particle production. In general they depend on
various factors such as spin of the emitted particles, whether the particle is localized on brane 
or can propagate in the bulk. The greybody factor can be computed by finding the absorption
cross section for the type of particle, in question, incident on the line element describing the adequate
black hole. It can be done in such a way because of the fact that Hawking's formula for the emission
rate for an outgoing particle at energy $\om$ equals the absorption cross section
for the same type of particle incoming at energy $\om$. Moreover,
outgoing transmission and ingoing absorption coefficients are equal. 
Therefore equilibrium still takes place if the black hole is located in a heat bath.
\par
In this section we shall focus our attention on finding the absorption probability in the low-energy regime.
First, our main task will be to derive an analytical expression of it by using the adequate approximate method.
In what follows,
we shall study massive scalar and phantom field emission in the background of a static spherically symmetric $(4 + n)$-dimensional
tense brane black hole.
The line element of such a black hole is subject to the relation
\be
ds^2 = - f(r)dt^2 + {dr^2 \over f(r)} + r^2 d\Omega^2_{n + 2},
\ee
where $f = 1 - \bigg( r_{0} / r \bigg)^{n + 1}$, $r_{0}$ is the radius of the black hole event horizon,
while $d\Omega^2_{n + 2}$ is a line element
on $S^{n + 2}$ sphere provided by the relation
\be
d\Omega^2_{n + 2} = d\theta^2 + \sum_{i = 2}^{n + 2} \prod_{j =1 }^{i -1} \sin^2 \phi_{j} d\phi_{i}^2.
\ee
In our considerations we take into account $S^{n + 2}$ sphere threaded by a codimension-2 brane, so
the range of the one of the angles, let us say, $\phi_{i}$ will be $ 0 \le \phi_{i} \le 2 \pi B$.
Parameter $B$ measures the deficit angle about axis parallel with the brane intersecting the sphere in question.
\par
Recent high precision observations have confirmed that our Universe undergoes a phase of acceleration expansion.
It happened that it is filled with {\it dark energy} which is responsible for the accelerated expansion.
Several models have been proposed in literature to explain this phenomenon (see \cite{cop06} and references therein).
In particular dynamical scalar fields such as {\it quintessence}, {\it k-essence}
and phantom field were taken into account.
In what follows we shall study the latter candidate for {\it dark energy} which is interesting because of its
negative kinetic energy. The Lagrangian for the phantom field is given by \cite{bro06,che09}
\be
L = - {1 \over 2} \na_{\mu}\tpsi~\na^{\mu}\tpsi - V(\tpsi),
\ee
where $V(\tpsi) = - {1 \over 2} m^2 \tpsi^2$.
In order to treat both phantom fields and ordinary massive scalar field one writes their
equations of motion in compact form. Namely,
\be
\na^{i}\na_{i}\tpsi - \eta~m^2 \tpsi = 0,
\ee
where $m$ is the mass of the adequate field,
$\eta = -1$ for the case of phantom fields and $\eta = 1$ for massive scalar field.
The scalars' field wave equation will be separable in this background if one uses the ansatz
\be
\tpsi (t, r, \theta, \phi_{i}) = e^{-i \om t}~R(r)~Y_{l}^{m}(\theta, \phi_{i}),
\ee
where $Y_{l}^{m}$ is a scalar spherical harmonics on the unit $(n + 2)$-sphere.
When $(n + 2)$-sphere is pierced by a topological defect one can introduce
{\it hyper-spherical} coordinates in the form:~
$0 < r < \infty,~0 \le \theta \le \pi,~-\pi \le \phi_{1} \le \pi,~\dots,~-\pi B \le \phi_{n + 1} \le \pi B$.
Having in mind results of Ref.\cite{coe02},
we may introduce $k = \sum_{i = 1}^{n + 1} c_{i}$, where $c_{i}$ are the integers separation constants.
Then, the multiple number $l$ has the form as
\be
l = k + \tc,
\ee
where $\tc = {c \over B}$ expresses the existence
of the aforementioned topological defect.
By virtue of the above ansatz one gets:
\be
{f(r) \over r^{n + 2}} {d \over dr} \bigg[
f(r) r^{n + 2} {dR \over dr} \bigg]
+ \bigg[ \om^2 - f(r) \bigg( \eta m^2 + {l (l + n + 1) \over r^2 }\bigg) \bigg]~R = 0.
\label{em1}
\ee
\par
Consequently, we shall obtain an analytical solution of the radial equation of motion by 
{\it approximation technique}. The main idea is to solve 
the radial equation of motion in the near horizon region and then
in the far-field limit. Next, one should smoothly match these two solutions in an intermediate region.\\
Let us begin with the near event horizon regime. Making a change of variables in Eq.(\ref{em1})
it may be rewritten as follows:
\be
f (1 - f) {d^2 R \over df^2} + (1 - f)~{dR \over df} + {1 \over (n + 1)^2}
\bigg[
{\om^2 r^2 \over f (1 - f)} - {(\eta m^2 r^2 + l (l + n + 1)) \over (n + 1)^2~(1 - f)}
\bigg] R = 0.
\label{a1}
\ee
We redefine $R(f) = f^{\alpha} (1 - f)^{\beta} F(f)$ and remove
singularities at $f = 0$ and $f = 1$. By means of this redefinition Eq.(\ref{a1})
can be transform to the of hypergeometric one, which yields
\be
f~(1 - f)~{d^2 F \over df^2} + [c - (1 + a + b)f]~{dF \over df} - ab F = 0.
\ee
It can be checked that the hypergeometric equation parameters satisfy
$a = b = \alpha + \beta$ and $c = 1 + 2 \alpha$. On the other hand, $\alpha$ and $\beta$
imply
\ben
\alpha_{\pm} &=& \pm {i \om r_{0} \over n + 1}, \\ \nonumber
\beta_{\pm} &=& {1 \over 2} \pm {1 \over n + 1} \sqrt{ \bigg(
l + {n + 1 \over 2} \bigg)^2 - (\om^2 - \eta m^2)~ r_{0}^2 }.
\een
Having in mind the criterion for the hyperbolic function to be convergent, i.e.,
$Re( c - a -b) > 0$ one has to select $\beta = \beta_{-}$. Just the general solution
of Eq.(\ref{a1}) may be written in the form
\be
R_{NH}(f) = A_{-}~f^{\alpha}~(1 - f)^{\beta}~F(\alpha + \beta, \alpha + \beta, 1 + 2 \alpha; f)
+ A_{+}~f^{- \alpha}~(1 - f)^{\beta}~F(\beta - \alpha, \beta - \alpha, 1 - 2 \alpha; f),
\ee
where $A_{\pm}$ are arbitrary constants. Because of the fact that no outgoing mode exists near
the event horizon of the considered black hole, we take $\alpha = \alpha_{-}$ and
put $A_{+}$ equal to zero. This leads to the following solution of equations of motion:
\be    
R_{NH}(f) = A_{-}~f^{\alpha}~(1 - f)^{\beta}~F(\alpha + \beta, \alpha + \beta, 1 + 2 \alpha; f).
\label{rr1}
\ee
Our next task is to match smoothly the near horizon solution $R_{NH}$ with the far field
one in the intermediate zone. To do this, first we change the expression of the hypergeometric
function near horizon zone from $f$ to $(1- f)$ by the standard relation (see \cite{abr66} relation $15.3.6$).
\par
Taking the limit $r \rightarrow \infty$ (or $f \rightarrow 1$) 
we are left with the Bessel kind of equation which implies
the general solution of the form
\be
R_{FF}(f) = {B_{+} \over r^{n + 1 \over 2}} J_{l + {n + 1 \over 2}}(\tom r)
+ {B_{-} \over r^{n + 1 \over 2}} Y_{l + {n + 1 \over 2}}(\tom r),
\label{rr2}
\ee
where $J_{\mu}$ and $Y_{\mu}$ are the Bessel function
of first and second order, respectively. $B_{\pm}$
are the integration constants. By $\tom$ we have denoted
$\tom = \sqrt{\om^2 - \eta m^2}$.
After having derived the analytical solutions of radial equation of motion for the considered fields
in the near horizon regime, Eq.(\ref{rr1}) and in the far field zone, Eq.(\ref{rr2}), we have to proceed
to build a solution valid at the whole radial regime. In order to do this we connect these two solutions in the intermediate region.
Thus, one has to stretch the near horizon solution to large values of $r$, while the far field one should be stretch towards 
small values of the $r$-coordinate.
On expanding in the limit $\tom r \rightarrow 0$ Eq.(\ref{rr2}), we conclude that 
\be
R_{FF}(\tom r \rightarrow 0) = {B_{+} ~r^l \over \Gamma (l + {n + 2 \over 3})}
\bigg(
{\tom \over 2} \bigg)^{l + {n + 1 \over 2}}
- {B_{-} \over r^{l + n + 1}}
\bigg(
{2 \over \tom} \bigg)^{l + {n + 1 \over 2}} {\Gamma (l + {n + 1 \over 2}) \over \pi},
\ee
while in the case of the near horizon solution $R_{NH}$ it yields
\be
R_{NH}(f) \cong A_{-} \bigg[
\bigg( {r \over r_{0}} \bigg)^l
{\Gamma( 1 + 2 \alpha)~ \Gamma(1 - 2 \beta) \over \Gamma(1 + \alpha - \beta)^2}
\bigg] +
A_{-} \bigg[
\bigg( {r_{0} \over r} \bigg)^{l + n + 1}~
{\Gamma( 1 + 2 \alpha)~ \Gamma(2 \beta - 1) \over \Gamma(\alpha + \beta)^2}
\bigg].
\ee
\par
\par
After a smooth matching was achieved, the ratio of the integration constants are provided by
the following:
\be
\cB = {B_{+} \over B_{-}} = - \bigg( {2 \over \tom r_{0}} \bigg)^{2l + n + 1}~
{\Gamma(l + {n + 1 \over 2})^2~ (l + {n + 1 \over 2})~\Gamma(1 - 2 \beta)~\Gamma( \alpha + \beta)^2
\over
\pi~\Gamma(1 + \alpha - \beta)^2~ \Gamma(2 \beta -1)}.
\label{coef}
\ee
After having calculated the ratio $\cB$ the absorption probability may be written in the form
\be
\mid \cA \mid^2 = {2i (\cB^{\ast} - \cB) \over \cB \cB^{\ast} + i (\cB^{\ast} - \cB) + 1}.
\ee
It turns out that the above relation can be simplified due to the fact that we are using the low-energy limit.
In this case $\cB \cB^{\ast} \gg i (\cB^{\ast} - \cB) \gg 1$. Consequently,
under this assumptions we conclude that
\be
\mid \cA \mid^2 \simeq 2 i \bigg( {1 \over \cB} - {1 \over \cB^{\ast}} \bigg).
\ee
Having in mind Eq.(\ref{coef}), one finds that the absorption probability can be brought to the following form:
\be
\mid \cA \mid^2 =
{4 \pi^2 ~\om ~\tom^{2l + n + 1} \over 2^{4l \over n +1}}~
\bigg({r_{0} \over  2} \bigg)^{2l + n + 2}~
{\Gamma(1 + {l \over n + 1})^2 \over
\Gamma({1 \over 2} + {l \over n + 1})^2~ \Gamma( l + {n + 3 \over 2})^2}.
\ee
It can be seen that $\mid \cA \mid^2$ does depend on deficit angle parameter $B$ through $l$.\\
We complete this section by finding 
the luminosity of the Hawking radiation for massive scalar and phantom field.
Let us turn to study the luminosity of the Hawking radiation for the mode $l = 0$ which
plays the dominant role in the greybody factor. It is provided by
\ben
L &=& \int_{0}^{\infty} {d \om \over 2 \pi}~\mid \cA \mid_{l = 0}^2 {\om \over e^{\om \over T_{BH}} - 1} \\ \nonumber
&=& {2 ( {r_{0} \over 2} )^{n + 2} \over \Gamma( {n + 3 \over 2})^2}~
T_{BH}^{n + 2}
\bigg[
T_{BH}^{2}~\Gamma(n + 4)~\zeta (n + 4) - {\eta m^2 \over 2}~(n + 1)~\Gamma(n + 2)~\zeta (n + 2)
\bigg],
\een
where $T_{BH} = {(n + 1) \over 4 \pi r_{0}}$ is the temperature of the tense brane black hole 
while $\zeta$ is Riemann zeta function.

\par
In order to obtain the absorption cross section of a plane wave incident
on the $(4 + n)$-dimensional tense brane black hole we expand a plane wave as follows \cite{das9697}:
\be
e^{i \tom r} \rightarrow {K~e^{- i \tom r} \over r^{n + 4 \over 2}} Y_{00} + other~ terms~ (higher~ multipole~ moments),
\label{kkk}
\ee
where $Y_{00} = {\Omega_{(2 + n)}}^{-1/2}$  is the normalized $s$-wave function on $(n + 2)$-dimensional sphere.
\par
As one can see, we take into account only the radially inward momentum component and ignore modes without spherical
symmetry. One can determine $K$ by integrating both sides of Eq.(\ref{kkk}) over $(n + 2)$-dimensional sphere
with deficit angle. Then, looking only at far region zone where $\tom~r \gg 1$, one extracts the ingoing modes. 
It can be proved that $K$ satisfies the following:
\be
\mid K \mid^2 =
{1 \over  \tom^{n + 2}}~{\Gamma( {n + 3 \over 2}) \over  \pi^{n + 3 \over 2}}~
4~ B~ \pi^{n + 2}~ 2^{n + 1}.
\label{kk}
\ee
Applying relation (\ref{kk}), we arrive to the relation for low-energy cross section. Thus,
it is provided by the following equation for the case when
$l = 0$:
\be
\sigma_{0} = \mid \cA \mid^2 \mid K \mid^2 =
\bigg( {\om \over \tom} \bigg)~ {2 B \pi^{n + 3 \over 2} \over \Gamma( {n + 3 \over 2})}~(r_{0})^{n + 2}
= \bigg( 1 - {\eta m^2 \over \om^2} \bigg)^{- {1 \over 2} }~\tA_{H},
\label{ss}
\ee 
where $\tA_{H} = {2 B \pi^{n + 3 \over 2} \over \Gamma( {n + 3 \over 2})}~(r_{0})^{n + 2}$
is the area of the event horizon of the considered black hole.
If further on, one assumes that the mass of the fields in question is small comparing to $\om$ the first coefficient 
in Eq.(\ref{ss}) is equal to one. Thus, at the massless limit of scalar field the absorption
cross section for $s$-wave coincides with the area of the black hole event horizon pierced by a topological defect.
One should remark that $\sigma_{0}$ has the correct limit in four-dimensions as well as for the massless limit.

\section{The Absorption Probability and Hawking Radiation in the Spacetime of a Tense Brane Black Hole}

In this section we present plots of the absorption probability and luminosity of Hawking radiation
as a function of $\om$ for different parameters. Namely, in Fig.1 we plotted $\mid \cA \mid^2$
for massive scalar and phantom field with respect to $\om$ for different spacetime dimension $n$. We take
$n = 1, \dots 4$, respectively. Other calculation parameters are:~$k = 0,~c = 1,~B = 0.9,~\tM = 1,~m = 0.05.$
$\tM$ has the form as $r_{0} = \tM / B^{1 \over n + 1}$.
One can observe that the higher dimension we take into account the smaller absorption
probability we obtain. In Fig.2 we depict the dependence of $\mid \cA \mid^2$ on frequency $\om$
for various $k$. The rest of the calculation parameters are the same as in Fig.1. From left
to right we plotted this dependence for five and six-dimensional tense brane black hole.
We also studied changes of the absorption probability for the considered fields for various values
of parameter $c$, respectively for $n = 1$ and $n =2 $ tense brane black hole. Calculation parameters we took are:
~$k = 0,~B = 0.9,~\tM = 1,$ and mass of the both fields is equal to $ 0.05$ (see Fig.3).\\
Fig.4 is devoted to the dependence of $\mid \cA \mid^2$ versus $\om$ for different mass of the scalar 
field in question, both for five and six-dimensional tense brane black hole (panel from left to right).
We considered $m = 0.01,~0.1$ and $0.15$. In Fig.5 the same was done for  phantom field.
One can observe that the absorption probability for massive scalar field decreases as mass of the field
increases, however for the phantom field $\mid \cA \mid^2$ increases with the increase of the mass parameter.
The same behaviour was observed studying the bulk absorption probability for scalars \cite{alb09}.
\par
Fig.6 depicts $\mid \cA \mid^2$ for different values of $\tM = 1.2,~1,~0.8$ for $n = 1$ and $n =2 $, respectively. 
In this plot the calculation parameters are: ~$k = 0,~c = 1,~B = 0.9,~m = 0.05$. One can observe that the absorption probability
decreases as $\tM$ decreases for both considered fields. 
The decrease of the absorption probability is greater for massive scalar field.
In Fig.7 we plotted absorption probability for massive scalar and phantom field
for different parameter $B = 1,~0.9, ~0.8$, which is connected with deficit angle. 
Calculation parameters are:~$k = 0,~c = 1,~\tM = 1,~m = 0.05$. We consider five and six-dimensional tense brane black hole
(panels from left to right, respectively). One has the situation
that the smaller $B$ is the smaller $\mid \cA \mid^2$ we achieve. One should recall that $B$ is         
connected with the brane tension in such a way that the smaller $B$ we have the greater tension is exerted
on the black hole. We observe the same behaviour for both fields in question. But it turned out that the brane tension has
less influence on phantom field than on massive scalar one (curves describing behaviour of phantom field always lie over
massive scalar ones).
\par
Fig.8 is devoted to the luminosity of Hawking radiation $L$ of massive scalar and phantom particles 
propagated in the considered background
with respect to mass of the fields in question, for various spacetime dimensions $n = 1, \dots 4$.
The other calculation parameters are: $B = 0.9$ and $\tM = 1$.

\section{Conclusions}
The future experiments which are planned to be conducted at LHC in the range of TeV gravity
will offer us possible window to explore the presence of extra dimensions. They will also help us to verify
brane worlds concepts. In this paper we studied Hawking emission of massive scalar and phantom fields in the
spacetime of $(4 + n)$-dimensional tense brane black hole. We have elaborated analytically
Eqs. of motion for the aforementioned degrees of freedom and by means of {\it matching technique}
one finds an analytical expression for the absorption cross section, luminosity of Hawking radiation
and the low-energy cross section. Our analytical considerations are supplemented by plots
of $\mid \cA \mid^2$ and the luminosity of Hawking radiation $L$. It turned out that
$\mid \cA \mid^2$ decreased with the number of extra dimensions $n$ for fields in question. The finite brane tension
also suppresses the emission rate of the considered fields in the spacetime under consideration.
\par
It will be not amiss to study fermion degrees of freedom in the spacetime in question.
We hope to return to this problem elsewhere.

 

\begin{acknowledgments}
This work was partially financed by the Polish budget funds in 2009 year as
the research project.
\end{acknowledgments}


\pagebreak

\begin{figure}
\begin{center}
  \includegraphics[width=0.85\textwidth]{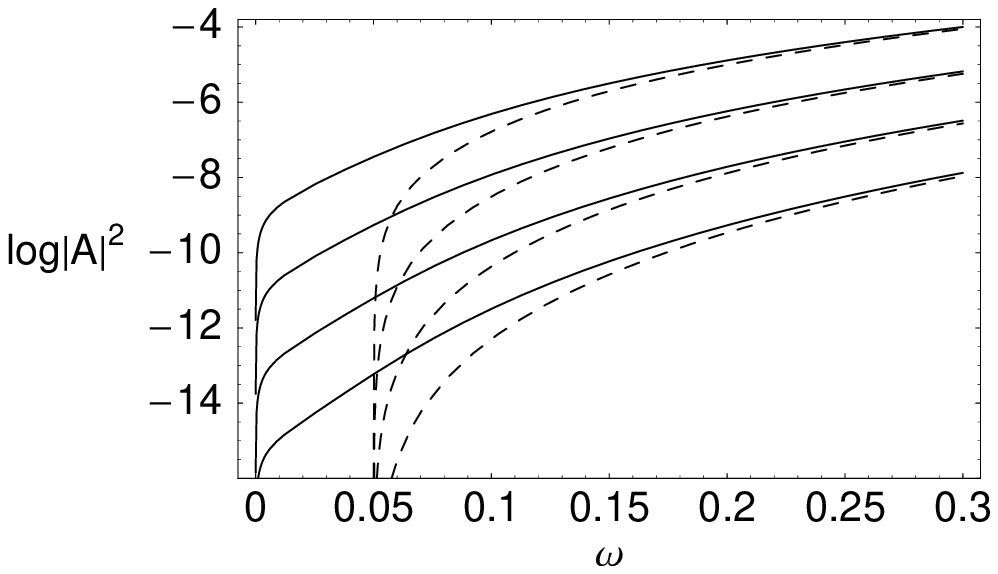}
\end{center}
\caption{Absorption probability $\mid \cA \mid^2$ for massive scalar field and phantom particles as a function of
$\om$, for different space dimensionality $n = 1,~2,~3,~4$ (curves from the top to the bottom, respectively).
Calculation parameters are:~$c =1,~k = 0,~\tM = 1,~B = 0.9,~m = 0.05$. Continuous line depicts phantom field, while
dashed one massive scalar field.}
\label{fig:fig1}
\end{figure}

\pagebreak

\begin{figure}
\begin{center}
  \includegraphics[width=0.85\textwidth]{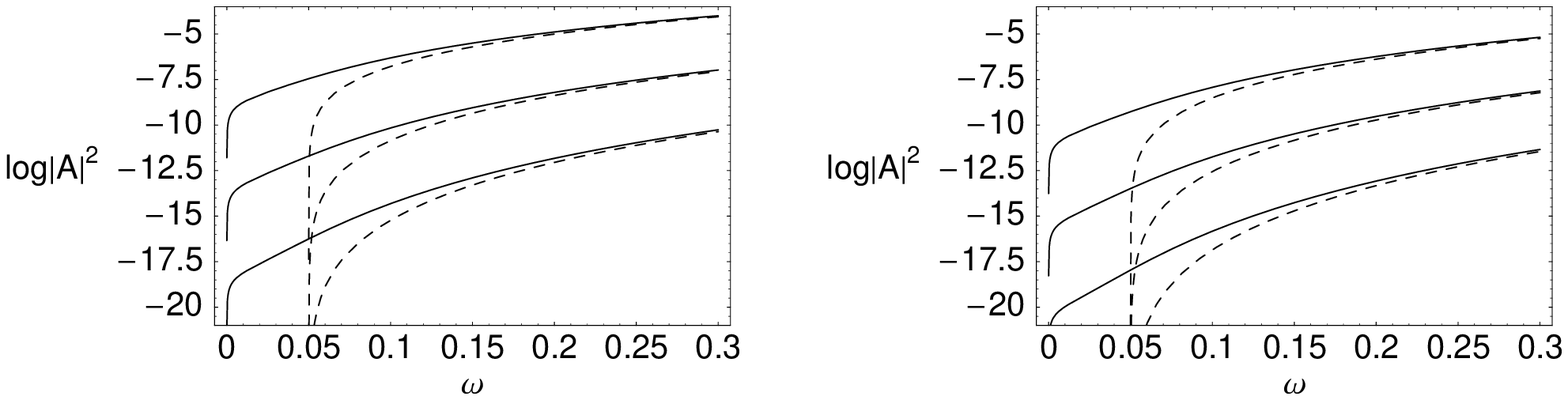}
\end{center}
\caption{Absorption probability $\mid \cA \mid^2$ for massive scalar and phantom fields as a function of
$\om$, for different values of $k = 0,~1,~2$ (curves from the top to the bottom, respectively)
and for five and six-dimensional tense brane black hole (left and right panel, respectively).
The rest of the calculation parametrs as in Fig.1. Dashed line depicts massive scalar field, while 
continuous one phantom field.}
\label{fig:fig2}
\end{figure}

\pagebreak

\begin{figure}
\begin{center}
  \includegraphics[width=0.85\textwidth]{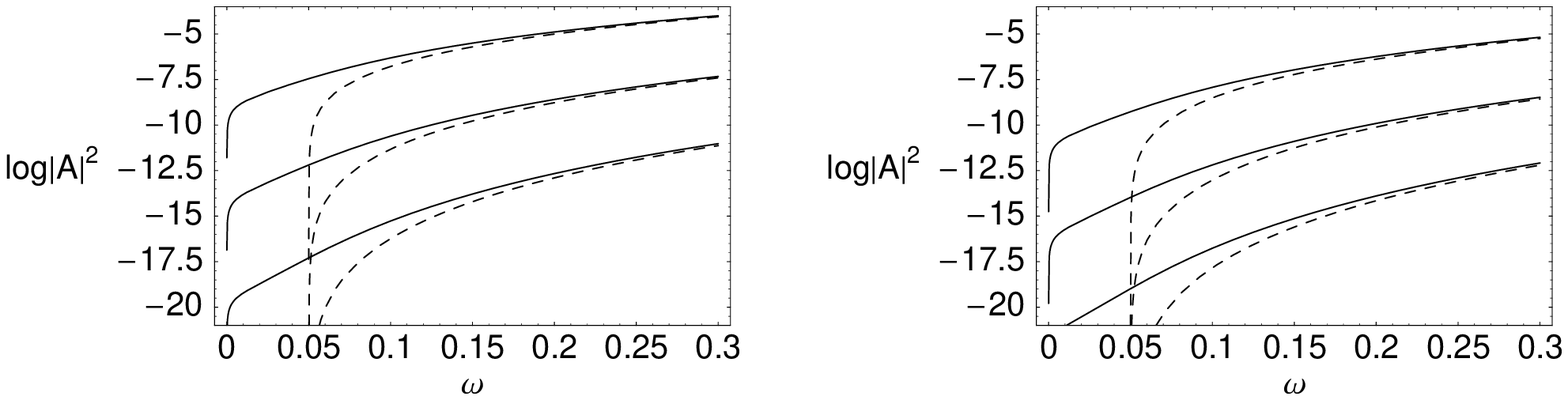}
\end{center}
\caption{Absorption probability $\mid \cA \mid^2$ for massive scalar (dashed line) and phantom (continuous line)
particles in a spacetime of five and six-dimensional tense brane black hole (left and right panel, respectively),
for different values of $c = 1,~2,~3$ (curves from the top to the bottom, respectively).
Calculation parameters are:~$k = 0,~\tM = 1,~B = 0.9,~m = 0.05$ and $n = 1,~ 2$.}
\label{fig:fig3}
\end{figure}

\pagebreak

\begin{figure}
\begin{center}
  \includegraphics[width=0.85\textwidth]{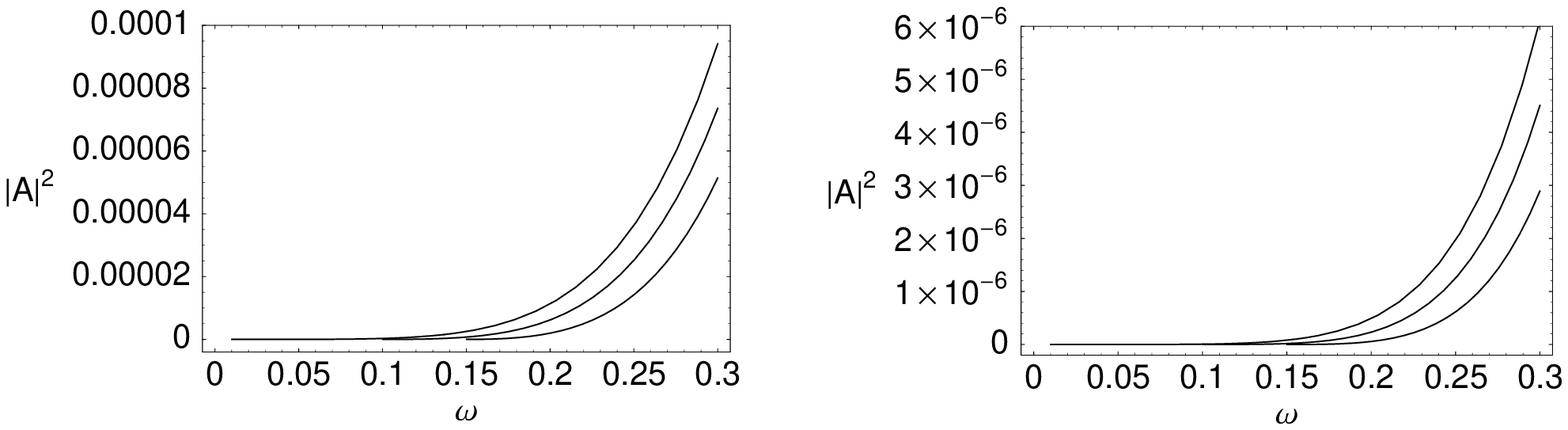}
\end{center}
\caption{Absorption probability $\mid \cA \mid^2$ for massive scalar
particles in a spacetime of five and six-dimensional tense brane black hole (left and right panel, respectively).
Curves from the top to the bottom are for different values of mass of the field in question
$m = 0.01,~0.1,~0.15$.
Calculation parameters are:~$c =1,~k = 0,~\tM = 1,~B = 0.9$ and $n = 1,~ 2$.}
\label{fig:fig4}
\end{figure}
\pagebreak

\begin{figure}
\begin{center}
  \includegraphics[width=0.85\textwidth]{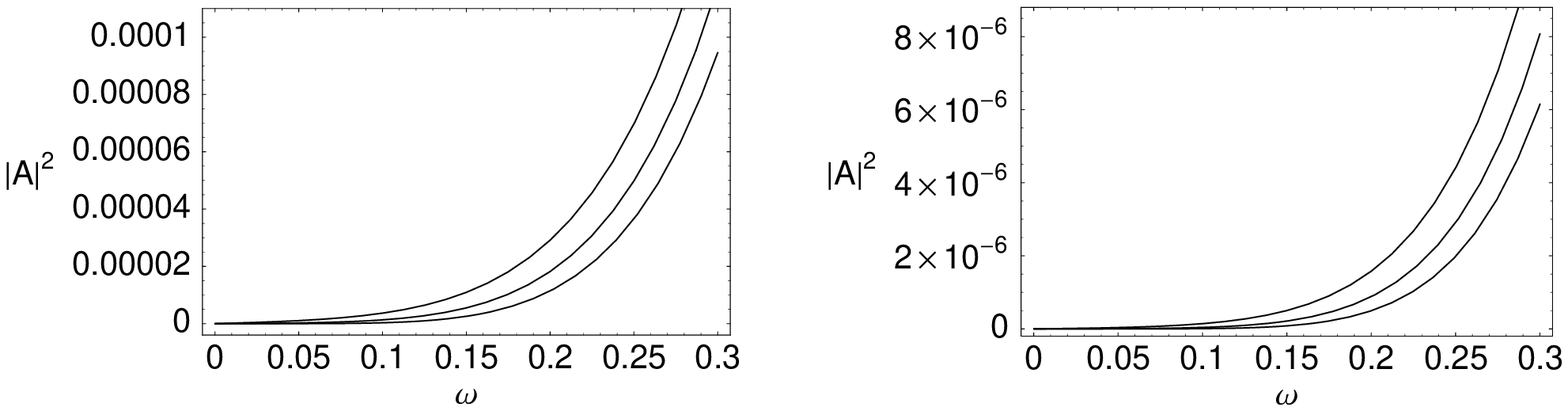}
\end{center}
\caption{Absorption probability $\mid \cA \mid^2$ for phantom
particles in a spacetime of five and six-dimensional tense brane black hole (left and right panel, respectively).
Curves from the bottom to the top are for different values of mass of the field in question
$m = 0.01,~0.1,~0.15$.
Calculation parameters are:~$c =1,~k = 0,~\tM = 1,~B = 0.9$ and $n = 1,~ 2$.}
\label{fig:fig5}
\end{figure}

\pagebreak

\begin{figure}
\begin{center}
  \includegraphics[width=0.85\textwidth]{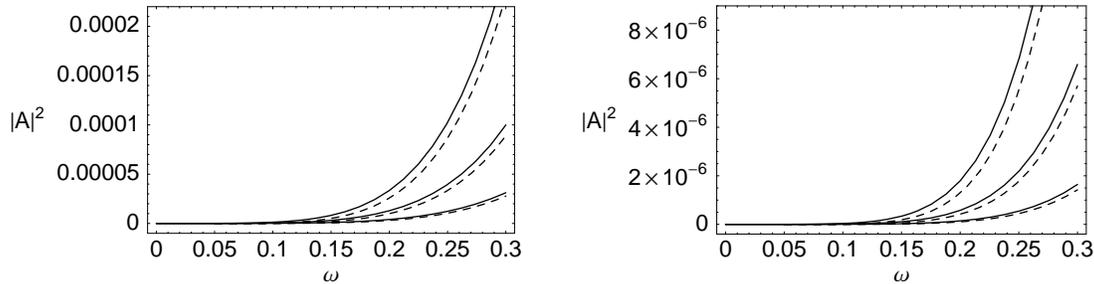}
\end{center}
\caption{Absorption probability $\mid \cA \mid^2$ versus $\om$ for 
different values of $\tM = 1.2,~1,~0.8$ (curves from the top to the bottom) and
for different spacetime dimensionality $n = 1,~2$ (panels from left to the right).
Other calculation parameters as in Fig.1. Dashed line depicts massive scalar field, while 
continuous one phantom field.}
\label{fig:fig6}
\end{figure}

\pagebreak

\begin{figure}
\begin{center}
  \includegraphics[width=0.85\textwidth]{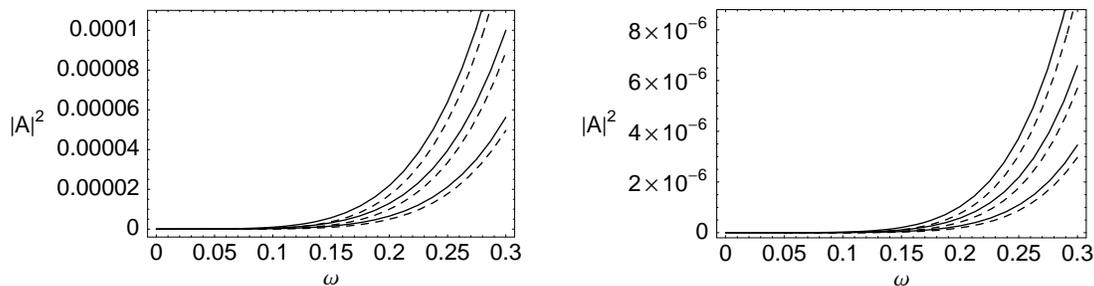}
\end{center}
\caption{Absorption probability $\mid \cA \mid^2$ versus $\om$ for 
scalar particles (dashed line) and phantom field (continuous line), for different 
values of $B = 1,~0.9,~0.8$ (curves from the top to the bottom, respectively) 
and
for different spacetime dimensionality $n = 1,~2$ (panels from left to the right).
Other calculation parameters as in Fig.1.}
\label{fig:fig7}
\end{figure}

\pagebreak

\begin{figure}
\begin{center}
  \includegraphics[width=0.85\textwidth]{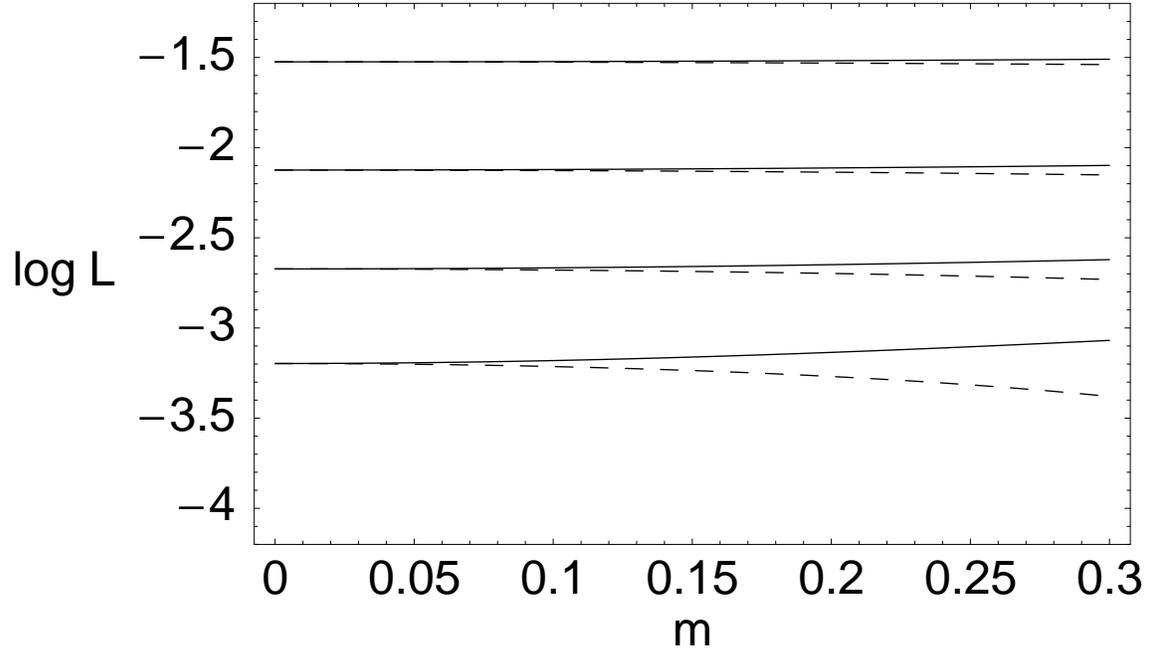}
\end{center}
\caption{The luminosity of Hawking radiation $L$ versus $m$ propagating on a tense brane black hole spacetime
for different space dimensionality $n = 1,~2,~3,~4$ (curves from the bottom to the top, respectively).
Other calculation parameters: $B = 0.9, ~\tM = 1$. Continuous line depicts phantom field, while 
dashed one massive scalar particles.}
\label{fig:fig8}
\end{figure}

\end{document}